\documentclass[aps,pre,reprint,twocolumn,superscriptaddress]{revtex4-2}

\usepackage{graphicx}
\usepackage{epstopdf, epsfig}
\usepackage{hyperref}
\usepackage{amsmath,amsfonts,amssymb}
\usepackage{subfigure}

\usepackage{graphicx}	
\usepackage{dcolumn}	
\usepackage{bm}			

\usepackage[utf8]{inputenc}
\usepackage{mathptmx}
\usepackage{etoolbox}
\usepackage{lipsum}
\usepackage{xcolor}
\usepackage[normalem]{ulem}
\usepackage{cancel}

\begin{document}

\title[Thermo-mechanic invariant]{A conserved thermo-mechanic invariant in extended fluid description of collisionless plasmas}

\author{E. S. Uchava}
\affiliation{Evgeni Kharadze Georgian National Astrophysical Observatory, Abastumani, 0301, Georgia}
\affiliation{Nodia Institute of Geophysics, Tbilisi State University, Tbilisi, Georgia}

\author{A. G. Tevzadze}
\affiliation{Evgeni Kharadze Georgian National Astrophysical Observatory, Abastumani, 0301, Georgia}
\affiliation{Ivane Javakhishvili Tbilisi State University, Tbilisi, 0179, Georgia}

\begin{abstract}

We investigate linear perturbations of an incompressible, weakly collisional, anisotropic plasma 
in the low frequency limit using an extended 16-moment fluid description that retains parallel 
and perpendicular heat fluxes. We identify a new class of linear perturbations associated with a 
conserved thermo-mechanic invariant, a time independent, aperiodic structure involving coupled perturbations of heat fluxes,  velocity, and magnetic field. In the standard CGL limit, where heat fluxes are neglected, no direct analogue of this invariant exists. Retaining heat flux dynamics alters the  linear structure of the system promoting third order velocity moments to autonomous variables and  gives rise to a thermo-mechanic mode with coupled thermal, kinetic, and magnetic components.  The associated perturbations are inherently localized, favoring compact, 
filamentary aperiodic structures. The thermo-mechanic invariant reveals a previously unexplored stationary sector of collisionless anisotropic plasma dynamics, characterized by a fixed algebraic polarization that enforces time 
independent relations among the relevant perturbation fields.

\end{abstract}


\maketitle

\section{Introduction}

A weakly collisional, rarefied plasma may exhibit anisotropic behavior under the influence 
of the Lorentz force resulting from an external magnetic field. In ionized media, the particle 
gyration radius around the magnetic field lines can be shorter than the mean free path of particle
collisions, resulting in directional anisotropy, whereby the effective pressure and temperature 
differ along and perpendicular to the magnetic field. Understanding the behavior 
of weakly collisional, anisotropic plasmas is essential for interpreting a wide range 
of astrophysical and laboratory systems, where magnetic geometry and heat-flux dynamics play 
a central role in energy transport, stability, and long-time evolution. Such conditions are 
encountered in laboratory plasma experiments, the solar wind, stellar and galactic outflows, 
and many other large-scale ionized flows throughout the Universe.

A comprehensive mathematical description of anisotropic plasmas is given by the 
Boltzmann–Vlasov kinetic formulation, which governs the evolution of the particle 
distribution function in phase space. While formally complete, its complexity often obscures
physical interpretation and limits practical applicability. Reduced fluid models therefore play 
an important role in describing collective plasma behavior: by taking velocity moments of the 
kinetic equation, they yield a hierarchy of coupled fluid equations that capture large-scale 
plasma dynamics in a more tractable form. From a practical standpoint, the formally infinite 
hierarchy must therefore be closed through a specific physical approximation. 
As a classical example, ideal magnetohydrodynamics (MHD) corresponds to a lowest-order 
fluid truncation retaining five kinetic moments (density, momentum, and isotropic pressure), 
coupled to the three components of the magnetic field.

A widely used extension of MHD to weakly collisional, magnetized plasmas is described by the 
Chew–Goldberger–Low (CGL) theory \cite{CGL}, with a recent review and discussion given in
\cite{Hunana2019}. In collisionless plasmas, reduced fluid descriptions can be systematically 
organized according to the kinetic information retained, as determined by both the level of moment truncation and the underlying physical assumptions. 
Within the usual moment counting framework, the CGL formulation corresponds to 
an anisotropic fluid model, obtained from a drift-kinetic reduction of the Vlasov equation under 
the assumptions of double adiabatic invariance and vanishing heat fluxes.
The resulting system yields a self-consistent description with anisotropic pressure components
parallel and perpendicular to the magnetic field. Owing to its simplicity and clear physical
interpretation, it remains widely employed and is arguably one of the most successful fluid 
approximation for anisotropic collisionless plasmas. Nevertheless, by construction, 
the CGL formulation omits dynamical heat flux effects that can play an important role in 
low frequency and aperiodic plasma evolution.

Building on the CGL description, a natural next step is to generalize the fluid framework by 
extending the velocity moment hierarchy in a systematic and conserving manner 
\cite{Hammett92,Mahajan08}. 
Grad-type closures provide arguably the most natural approach for such a generalization, 
as they truncate the Boltzmann–Vlasov moment hierarchy at a finite order while preserving 
the fundamental conservation laws \cite{Grad1949}. In this approach, the
distribution function is expanded in a finite set of orthogonal velocity moments, and closure 
is achieved by relating higher order moments to the retained lower order moments. When applied 
to weakly collisional or collisionless magnetized plasmas, this procedure yields a 
self-consistent fluid model in which mass, momentum, and energy are conserved by construction, 
provided no ad hoc dissipative terms are introduced. 

An important extension of CGL is the 16-moment Grad-type formulation, which retains anisotropic 
pressure components together with their associated heat fluxes as independent dynamical variables,
resulting in a closed fluid system. Such models can be derived following two complementary
routes. One proceeds directly from the classical Grad expansion of the Boltzmann–Vlasov equation, 
as developed for collisionless magnetized plasmas \cite{Oraevskii68,Dzhalilov08,Kuznetsov10}, 
where closure is imposed at the level of fourth-order velocity moments. 
The other starts from a gyrokinetic or drift-kinetic ordering and constructs a gyrofluid system 
by taking moments of the reduced kinetic equation \cite{Ramos03}, leading to an equivalent 
16-moment structure with heat fluxes. 
Despite their conceptually distinct theoretical constructions, these two formulations are
physically equivalent in the collisionless, low-frequency, long-wavelength regime, where
finite-Larmor-radius effects are negligible, since both elevate the parallel and perpendicular
heat fluxes to independent dynamical variables with their own evolution equations, thereby
retaining the same set of moments and conservation properties.
As a result, the 16-moment model captures 
essential heat flux driven thermo-mechanical couplings while remaining applicable at frequencies 
well below the cyclotron frequency and at length scales large compared to the gyroradius. 
In the following, we refer to this formulation as a 16-moment extended fluid model, emphasizing 
the retained dynamical moments rather than the particular closure construction.
\textcolor{black}{The specific form of the heat-flux evolution equations used here corresponds to the standard Grad-type 16-moment closure for collisionless plasmas (see, e.g., \cite{Oraevskii68,Dzhalilov08}), where third-order moments are retained as independent dynamical variables.}

An intermediate class of reduced descriptions is provided by gyrofluid models, which retain a fluid formulation while being derived systematically from gyrokinetic or drift-kinetic theory through a controlled separation between macroscopic plasma scales and the particle gyroradius.
Reduced kinetic and gyrofluid formulations, together with related extended anisotropic fluid models, have been developed in \cite{Hazeltine85,Ilgisonis90,Brizard1992}.
In the long wavelength, low frequency limit, where finite Larmor radius effects are negligible,
gyrokinetic analyses show that pressure anisotropy emerges naturally and that gyrofluid formulations reduce to extended moment systems equivalent to 16-moment models, thereby providing a natural bridge between extended fluid descriptions and fully kinetic theory~\cite{Snyder97,Schekochihin05}.
In this limit, as noted above, gyrofluid models retain anisotropic pressures together with parallel and perpendicular heat fluxes as independent dynamical variables, yielding a 16-moment extended fluid system ~\cite{Ramos03}.

A further refinement of reduced fluid modeling is provided by Landau-fluid formulations, 
which complement local fluid closures by incorporating particle–wave resonances, most notably 
Landau damping. This is accomplished through nonlocal closures along the magnetic field that 
reproduce the linear kinetic response of collisionless plasmas and improve the quantitative 
accuracy of fluid descriptions in resonant regimes \cite{Hammett90,Snyder97}.
Because heat fluxes are prescribed through such nonlocal operators rather than evolved as 
independent dynamical variables, Landau-fluid models are not always optimally suited for 
low frequency or aperiodic plasma dynamics in which autonomous heat flux evolution plays 
an important role.

The choice of a fluid model depends on the physical effects and dynamical processes of interest, 
with different closures being valid in distinct asymptotic regimes. Because the selected closure 
can influence not only quantitative predictions but also the qualitative structure of the reduced 
dynamics, particularly at low frequencies, where non-dissipative thermo-mechanical couplings may 
persist, the 16-moment extended fluid model provides a well suited framework for collisionless, 
low frequency plasma dynamics. By retaining anisotropic pressures together with parallel and 
perpendicular heat fluxes, it constitutes the minimal extension of CGL theory that captures such 
couplings while remaining local in space and time. 
Unlike double adiabatic and related  invariant preserving closures, it allows heat flux perturbations to evolve as independent dynamical variables. 

The present paper examines the linear dynamics of aperiodic low-frequency perturbations in an 
anisotropic, magnetized plasma within the framework of an extended 16-moment fluid model that 
self-consistently retains pressure anisotropy and parallel and perpendicular heat fluxes. 
Rather than focusing on wave-like or normal mode solutions, we emphasize persistent aperiodic 
perturbation structures associated with heat flux dynamics. Using the linear perturbation 
analysis directly in real space coordinates, we identify a previously unrecognized time invariant conservative quantity, which we refer to as the thermo-mechanic invariant, and explicitly construct  the associated perturbation fields. 

The paper is organized as follows. Section II introduces the governing equations of the 
extended 16-moment anisotropic fluid model and derives the corresponding 
linearized perturbation equations appropriate to the aperiodic regime. 
Within this framework, the conditions under which the thermo-mechanic invariant emerges 
are examined in detail, and its dependence on heat flux dynamics and pressure anisotropy is 
clarified. Section III summarizes the main results and discusses their implications for the 
interpretation of linear plasma dynamics beyond a normal mode description. In particular, 
we comment on the possible role of persistent, aperiodic structures in shaping the long 
time evolution and global behavior of anisotropic plasma flows in a variety of physical contexts.

\section{Physical Model}

The dynamics of an incompressible, collisionless plasma with heat fluxes can be described 
within the framework of an extended 16-moment fluid model, 
as given by the following set of equations \cite{Uchava14,Uchava20}:
\begin{align}
\partial_t \mathbf{V} + (\mathbf{V}\cdot\nabla)\mathbf{V}
&= -\frac{1}{\rho}\nabla P_\perp
+ \frac{1}{4\pi\rho}\big[(\nabla\times\mathbf{B})\times\mathbf{B}\big]
\nonumber\\
&\quad + \frac{1}{\rho}
(\mathbf{B}\cdot\nabla)\!\left[
(P_\perp - P_\parallel)\frac{\mathbf{B}}{B^2}
\right],
\label{mhd1} \\[4pt]
\partial_t \mathbf{B}
&= \nabla\times(\mathbf{V}\times\mathbf{B}),
\label{mhd2} \\
\nabla\cdot\mathbf{B}
&= 0, \label{mhd3} \\
\nabla\cdot\mathbf{V}
&= 0.
\label{mhd4}
\end{align}
which are complemented by the double-adiabatic equations governing the parallel ($P_{\parallel}$) 
and perpendicular ($P_{\perp}$) pressure components with respect to the magnetic field:
\begin{align}
\frac{\mathrm{d}}{\mathrm{d}t}
\left( \frac{P_{\parallel} B^{2}}{\rho^{3}} \right)
&= -\frac{B^{3}}{\rho^{3}}
\left[
(\mathbf{h}\cdot\nabla)
\left( \frac{S_{\parallel}}{B} \right)
+ \frac{2S_{\perp}}{B^{2}}
(\mathbf{h}\cdot\nabla) B
\right] ~,
\label{mhd5} \\[6pt]
\frac{\mathrm{d}}{\mathrm{d}t}
\left( \frac{P_{\perp}}{\rho B} \right)
&= -\frac{B}{\rho}
(\mathbf{h}\cdot\nabla)
\left( \frac{S_{\perp}}{B^{2}} \right) ~,
\label{mhd6}
\end{align}
where $S_{\parallel}$ and $S_{\perp}$ represent the fluxes of parallel and perpendicular thermal energy, respectively, both directed along the magnetic field.
Here $\mathbf{h}=\mathbf{B}/B$ is the unit vector in the direction of the magnetic field, and
$$
{{\rm d} \over {\rm d} t} \equiv \partial_t + (\mathbf{V} \cdot
\nabla) ~.
$$
The closed set of the fluid model is obtained by supplementing the governing equations with 
the heat-flux evolution equations, as follows:
\begin{align}
\frac{\mathrm{d}}{\mathrm{d}t}
\left( \frac{S_{\parallel} B^{3}}{\rho^{4}} \right)
=&
-\,j\,\frac{3 P_{\parallel} B^{3}}{\rho^{4}}
(\mathbf{h}\cdot\nabla)
\left( \frac{P_{\parallel}}{\rho} \right),
\label{mhd7} \\[6pt]
\frac{\mathrm{d}}{\mathrm{d}t}
\left( \frac{S_{\perp}}{\rho^{2}} \right)
=&
-\,j\,\frac{P_{\parallel}}{\rho^{2}}
\Bigg[
(\mathbf{h}\cdot\nabla)
\left( \frac{P_{\perp}}{\rho} \right)
+ \frac{P_{\perp}}{\rho}
\frac{P_{\perp}-P_{\parallel}}{P_{\parallel} B}
(\mathbf{h}\cdot\nabla) B
\Bigg].
\label{mhd8}
\end{align}
Here the parameter $j$ is introduced to represent the zero heat flux limit of the system, 
thereby allowing the obtained results to be reduced to the CGL limit.
\begin{equation}
j \equiv
\begin{cases}
1, & \text{when } S_{\perp}\neq 0 \text{ or } S_{\parallel}\neq 0, \\
0, & \text{when } S_{\perp}=0 \text{ and } S_{\parallel}=0.
\end{cases}
\end{equation}
The system of Eqs.~(\ref{mhd1})–(\ref{mhd8}) constitutes a closed set governing the dynamics of 
a weakly collisional plasma within an extended anisotropic fluid framework that retains 
heat flux moments. The increased complexity relative to standard MHD reflects the generalized 
equation of state, defined by Eqs.~(\ref{mhd5})–(\ref{mhd8}), which promotes the parallel and 
perpendicular pressures and heat fluxes to independent dynamical variables, thereby introducing 
additional degrees of freedom.

\subsection{Equilibrium state}

For the equilibrium configuration, we assume a uniform plasma density ($\rho=\text{const}$) 
and a steady flow aligned with a uniform background magnetic field ($B_0=\text{const}$). 
For simplicity, the $x$ axis is chosen to be aligned with the direction of the magnetic field.
\begin{equation}
{\bf B}_0 = (B_0, 0, 0) ~, \nonumber
\end{equation}
and describe the plasma in the frame co-moving with the flow, where $\mathbf{V}_0=0$. 
The background anisotropic pressure and heat flux parameters are specified in terms of components 
parallel and perpendicular to the magnetic-field direction:
\begin{align*}
P_{0\parallel} &= \text{const}, \qquad
P_{0\perp} = \text{const}, \\
S_{0\parallel} &= \text{const}, \qquad
S_{0\perp} = \text{const}.
\end{align*}
where plasma anisotropy parameter can be calculated as:
\begin{equation}
\alpha = {P_{0\perp} \over P_{0\|}} ~.
\end{equation}
\textcolor{black}{We note that constant background heat fluxes are admissible in this formulation, since only their spatial gradients enter the governing equations, and these vanish in a homogeneous state. In this sense, the equilibrium can be viewed as an idealized or quasi-equilibrium configuration with prescribed higher order moments.}

To parametrize the problem we introduce the following well known physical parameters: 
\begin{eqnarray}
{\rm Alfven~speed:} ~~&& V_A^2 = {B_0^2 \over 4 \pi \rho} ~,~	\nonumber \\
{\rm Directional~sound~speeds:} ~~&& C_\|^2 =  {P_{0\|} \over \rho} ~,~	
C_\perp^2 = {P_{0\perp} \over \rho}	~, \nonumber \\
{\rm Directional~plasma~parameters:} ~~&& \beta_\| = {4 \pi P_{0\|} \over B_0^2} ~,~ \beta_\perp = {4 \pi P_{0\perp} \over B_0^2}  ~. \nonumber
\end{eqnarray}
For the compactness of the notations we may introduce the normalized heat flux parameters:
\begin{equation}
q_\| = {S_{0\|} \over P_{0\|} V_A} ~,~~~
q_\perp = {S_{0\perp} \over P_{0\perp} V_A} ~,
\end{equation}
\textcolor{black}{In general, heat flux and pressure anisotropies are independent in collisionless plasmas, and may differ when kinetic effects become important for example, under strong Landau damping, during the development of pressure anisotropy instabilities (firehose or mirror), or in the presence of large temperature gradients along the magnetic field. In this work, we adopt the simplifying assumption that the heat flux anisotropy follows the pressure anisotropy, i.e. $q_\parallel = q_\perp$. Physically, this choice isolates the role of heat flux dynamics without introducing an additional independent anisotropy parameter associated with third order moments, providing a minimal and controlled extension beyond the CGL limit.}

Hence, we introduce parameters associated with the kinetic firehose mode:
\begin{equation}
{\beta_{\rm F} \equiv \beta_\perp - \beta_\| +1} ~,
\end{equation}
and anisotropy induced difference between the perpendicular and parallel sound speeds:
\begin{equation}
{\beta_\Delta \equiv \beta_\perp - \beta_\| = (\alpha-1) \beta_\|  }
\end{equation}
It is important to note that both \textcolor{black}{$\beta_{\rm F}$} and \textcolor{black}{$\beta_{\Delta}$} may take either positive 
or negative values, depending on the underlying microscopic stability conditions, such as 
the presence of the firehose instability (see Refs.~\cite{Uchava20,Longmire56,Chandra58,Parker58,Schekochihin05,Kunz14}) 
or the nature of the plasma anisotropy (e.g., when $\alpha<1$).

\subsection{Linear Perturbations}

The dynamics of linear perturbations about the equilibrium state described above can be analyzed 
using a standard linearization procedure.
For the shortness of the notations we introduce a generalized state vector collecting the physical variables of the problem:
\begin{equation}
\Psi = \left(V_x,V_y,V_z,P_\|,P_\perp,S_\|,S_\perp,B_x, B_y,B_z \right) ~.
\end{equation}
Hence, we introduce generalized vectors representing the equilibrium ($\boldsymbol{\Psi}_0$) 
and perturbed (\textcolor{black}{$\boldsymbol{\Psi}^\prime$}) components of the physical variables:
\begin{equation}
\Psi = \Psi_0 + \Psi^\prime ~.
\end{equation}
With the equilibrium state defined by the physical model introduced above, we obtain:
\begin{equation}
\Psi_0 = \left(0 , 0 , 0 ,P_{0\|} ,P_{0\perp} ,S_{0\|} ,S_{0\perp} ,B_0 , 0 ,0 \right) ~,
\end{equation}
Thus, by introducing linear perturbations about the background state and neglecting nonlinear terms, 
we derive a system of linear differential equations describing the dynamics of perturbations 
in an anisotropic medium. To express this system in nondimensional form, we employ 
a generalized state vector composed of normalized physical variables.
\begin{equation}
\psi = \left(v_x , v_y , v_z , p_\| , p_\perp , s_\| , s_\perp , b_x , b_y , b_z \right) ~,
\end{equation} 
where
\begin{equation}
\begin{aligned}
\mathbf{v} \equiv \frac{\mathbf{V}'}{V_A} ~,&~~~  \mathbf{b} \equiv \frac{\mathbf{B}'}{B_0} ~, \\
\textcolor{black}{p_\| \equiv \frac{P_\|'}{P_{0 \|}} ~,} &~~~
\textcolor{black}{p_\perp \equiv \frac{P_\perp'}{P_{0 \perp}} ~, } \\
s_\| \equiv \frac{S_\|'}{P_{0\|} V_A} ~,&~~~ s_\perp \equiv \frac{S_\perp'}{P_{0\perp} V_A} ~,
\end{aligned}
\end{equation}
and introduce a characteristic length-scale of the problem $L$ and adopt a non-dimensional 
description by using:
\begin{equation}
\begin{aligned}
t &\to t\,L / V_A, \\
x,y,z &\to Lx,\,Ly,\,Lz .
\end{aligned}
\end{equation}

Hence, the dynamical equations can be written in non-dimensional matrix form as follows:
\begin{equation}
\partial_t \psi_i = -{ {\bf L}_{i\textcolor{black}{k}}}\psi_{\textcolor{black}{k}} ~,
\end{equation}
where ${\bf L}_{i\textcolor{black}{k}}$ is a tenth-order matrix operator defined by:
\begin{widetext}
\begin{equation}
\mathbf{L}_{i\textcolor{black}{k}} =
\begin{pmatrix}
 \cdot & \cdot & \cdot & \beta_\| \partial_x & \cdot & \cdot & \cdot & \beta_\Delta \partial_x & \cdot & \cdot \\
 \cdot & \cdot & \cdot & \cdot & \beta_\perp \partial_y & \cdot & \cdot & \partial_y & -\beta_{\rm F} \partial_x & \cdot \\
 \cdot & \cdot & \cdot & \cdot & \beta_\perp \partial_z & \cdot & \cdot & \partial_z & \cdot & -\beta_{\rm F} \partial_x \\
 3\partial_x & \partial_y & \partial_z & \cdot & \cdot & \partial_x & \cdot &
 (2\alpha-1) q_{\parallel} \partial_x & \cdot & \cdot \\
 \partial_x & 2\partial_y & 2\partial_z & \cdot & \cdot & \cdot & \partial_x &
 -2 q_{\perp} \partial_x & \cdot & \cdot \\
 4 q_{\parallel} \partial_x & q_{\parallel}\partial_y & q_{\parallel}\partial_z &
 3j \beta_{\|} \partial_x & \cdot & \cdot & \cdot & \cdot & \cdot & \cdot \\
 2 q_{\perp} \partial_x & 2 q_{\perp} \partial_y & 2 q_{\perp} \partial_z &
 \cdot & j \beta_{\|} \partial_x & \cdot & \cdot &
 j \beta_{\Delta} \partial_x & \cdot & \cdot \\
 \cdot & \partial_y & \partial_z & \cdot & \cdot & \cdot & \cdot & \cdot & \cdot & \cdot \\
 \cdot & -\partial_x & \cdot & \cdot & \cdot & \cdot & \cdot & \cdot & \cdot & \cdot \\
 \cdot & \cdot & -\partial_x & \cdot & \cdot & \cdot & \cdot & \cdot & \cdot & \cdot
\end{pmatrix} ~,
\end{equation}
\end{widetext}
where the dots denote zero matrix elements. Using the divergence-free conditions for the velocity 
and magnetic-field perturbations,
\begin{eqnarray}
{\partial_x}\textcolor{black}{v_x^\prime}+{\partial_y}\textcolor{black}{v_y^\prime}+{\partial_z}\textcolor{black}{v_z^\prime} = 0 ~, \\
{\partial_x} \textcolor{black}{b_x^\prime} + {\partial_y} \textcolor{black}{b_y^\prime} +{\partial_z} \textcolor{black}{b_z^\prime} = 0 ~, 
\end{eqnarray}
we obtain the following stationary relation between the parallel and perpendicular components 
of the pressure perturbations:
\begin{equation}
\beta_\|{\partial_{xx} \textcolor{black}{p_\|^\prime}} = 
- \beta_\perp \Delta_\perp \textcolor{black}{p_\perp^\prime}
- \Delta \textcolor{black}{b_x^\prime}  \textcolor{black}{- \, 2\beta_\Delta \partial_{xx} b_x' } ~,
\end{equation}
where the following differential operators are introduced:
\begin{align}
\Delta_{\perp} &\equiv \partial_{yy} + \partial_{zz} ~, \\
\Delta &\equiv \partial_{xx} + \Delta_{\perp} ~.
\end{align}
Hence, we may reduce the system by taking its derivative and substituting linear perturbations 
of $V_z^\prime$, $B_z^\prime$, and $P_\|^\prime$ using the above equations. 
Let us define the reduced perturbation vector as:
\begin{equation}
\Phi_i = \left(v_x, v_y, p_\perp, s_\|, s_\perp, b_x, b_y \right) ~.
\end{equation}
Now, the temporal dynamics of linear perturbations may be described by:
\begin{equation}
\partial_t \partial_x \Phi_i={ {\bf N}_{i\textcolor{black}{k}}}\Phi_{\textcolor{black}{k}}, 
\end{equation}
\textcolor{black}{Note that an additional $\partial_x$ operator is introduced to facilitate further reduction using Eq. (24).}
Hence, the seventh-order matrix operator ${\bf N}_{i\textcolor{black}{k}}$ can be given by:
\begin{widetext}
\begin{equation}
{\bf N}_{i\textcolor{black}{k}} =
\left(
\begin{array}{ccccccc}
\cdot & \cdot & \beta_\perp \Delta_\perp & \cdot & \cdot &
\Delta_\perp + \beta_F \partial_{xx} & \cdot \\
\cdot & \cdot & - \beta_\perp \partial_{xy} & \cdot & \cdot &
-\partial_{xy} & \beta_{\rm F} \partial_{xx} \\
\partial_{xx} & \cdot & \cdot & \cdot &
-\partial_{xx} & 2 \alpha q_\perp \partial_{xx} & \cdot \\
-3 q_\| \partial_{xx} & \cdot &
3 j \beta_\perp \Delta_\perp & \cdot & \cdot &
3 j (\Delta \textcolor{black}{+} 2 \beta_\Delta \partial_{xx}) & \cdot \\
\cdot & \cdot & - j \beta_\| \partial_{xx} & \cdot & \cdot &
j \beta_\Delta \partial_{xx} & \cdot \\
\partial_{xx} & \cdot & \cdot & \cdot & \cdot & \cdot & \cdot \\
\cdot & \partial_{xx} & \cdot & \cdot & \cdot & \cdot & \cdot
\end{array}
\right), \label{N_mat}
\end{equation}
\end{widetext}
The non-stationary solutions of the linear perturbation system for incompressible anisotropic 
flows have already been analyzed using a sixth-order dispersion equation \cite{Dzhalilov08}. 
That analysis addresses the wave-like and unstable, time dependent sector of the spectrum. 
\textcolor{black}{Here we use a real-space formulation to complement this analysis in the stationary case; this approach is better suited for spatially inhomogeneous perturbations, as it provides a local representation of the invariant, preserves the underlying operator structure, and avoids the nonlocal reconstruction inherent to Fourier-space methods.}
In this sense, the present study provides a complementary eigenfunction analysis of the full 
seventh-order matrix, which focuses on the aperiodic sector and shows that it admits a 
non-zero, non-trivial stationary solution.

\subsection{Stationary Solution}

The time invariant quantity of the system, if present, can be expressed as a linear combination of perturbations:
\begin{equation}
W({\bf r},t) \equiv {\bf g}_i \Phi_i({\bf r},t) ~,
\end{equation}
where ${\bf g}_i$ is vector generator of the invariant quantity $W$. 
\textcolor{black}{Since the present analysis is restricted to the linearized perturbation system, we seek first order invariants that are linear functionals of the perturbation vector, corresponding to conserved quantities associated with the underlying linear operator that governs the perturbation dynamics.}
The existence of a conservative quantity can be established through the presence of a nontrivial solution ($\Phi_i \not= 0$) of the following stationarity condition:
\begin{equation}
\partial_t W({\bf r},t) = 0 ~.
\end{equation}
Using Eq. (30) in the stationarity condition (31) and assuming that generator ${\bf g}_i$ is 
uniform both in space and time we may derive the following condition:
\begin{equation}
\partial_t \partial_x W = {\bf g}_i \partial_t \partial_x \Phi_i = 0 ~,
\end{equation}
\textcolor{black}{where using a homogeneous ${\bf g}_i$ ansatz is a natural choice in an equilibrium without an intrinsic spatial scale.}
Hence, Eq. (28) leads to the realizability condition for the existence of the aperiodic mode:
\begin{equation}
{\bf g}_i {\bf N}_{i\textcolor{black}{k}} \Phi_{\textcolor{black}{k}} = 0 ~.
\end{equation}
Following the \textcolor{black}{structural properties} of the matrix operator ${\bf N}_{i\textcolor{black}{k}}$, we may derive \textcolor{black}{(see Appendix A for details)}:
\begin{equation}
{\bf g}_i = \left(\begin{array}{ccccccc}
-3 j \left(\beta_\Delta \partial_{xx} + {\bf f} \right) \\
0 \\ 0 \\ {\bf f} \\
-3 \alpha \beta_\Delta \Delta_\perp \\
3 q_\| {\bf f} \\ 0 
\end{array} \right) ~, 
\label{gi}
\end{equation}
where the operator $\bf f$ is introduced for the shortness of the notations:
\begin{equation}
{\bf f} \equiv \Delta \, + \, \beta_\Delta \left( \partial_{xx} \textcolor{black}{+} \alpha \Delta_\perp \right) ~. 
\end{equation}
Hence, utilizing the vector generator derived from Eqs. (34-39), we can obtain the explicit form 
of the time invariant variable as follows:
\begin{eqnarray}
W({\bf r}) &=& {\bf f} \left( s_\|({\bf r},t)- 3 j v_x({\bf r},t) + 3 q_\| b_x({\bf r},t) \right) 
\nonumber \\ 
&& -  3 j \beta_\Delta \partial_{xx} v_x({\bf r},t)
-3 \alpha \beta_\Delta \Delta_\perp s_\perp({\bf r},t) ~.
\end{eqnarray}
It appears that the perturbations of the linear stationary mode $W$ encompass heat fluxes in both 
parallel and perpendicular directions, as well as perturbations in velocity and magnetic 
field along the background magnetic field. Consequently, we can identify 
$W$ as the \textbf{\textit{thermo-mechanic invariant}} of the linear system.
Existence of such conserved quantity highlights the presence of stable, non-oscillatory modes, 
which may play a significant role in the redistribution of energy within weakly collisional 
anisotropic plasmas.
\textcolor{black}{
It seems that the thermo-mechanic mode establishes a local algebraic constraint among the perturbation fields, enabling direct identification of the aperiodic component of the solution in real space. In this way, it provides a natural diagnostic for distinguishing stationary perturbations from propagating or unstable components of the linear dynamics.}

To recover the standard CGL limit, one sets
$j=0$, $q_\parallel=q_\perp=0$, and
$s_\parallel=s_\perp=0$. Under these conditions, the invariant
reduces identically to
\begin{equation}
W=0 ~,
\end{equation}
demonstrating that no nontrivial direct analogue of the present
invariant exists within the CGL model.
Nevertheless, it is useful to decompose the invariant as
\begin{equation}
W=W_S+W_{VB} ~,
\end{equation}
with the individual contributions given by
\begin{align}
W_{S} &= \left( \Delta + \beta_\Delta ( \partial_{xx} \textcolor{black}{+} \alpha \Delta_\perp )\right) s_\| - 3 \alpha \beta_\Delta \Delta_\perp s_\perp ~, \\
W_{VB} &= -3 \left[ (1 + 2 (\alpha-1)\beta_\|) \partial_{xx} 
+ (1 \textcolor{black}{+} (\alpha-1)\beta_\perp) \Delta_\perp \right] {v_x}  + \nonumber \\
& 3 \left[ (1 + (\alpha-1)\beta_\|) \partial_{xx} + (1 \textcolor{black}{+} (\alpha-1)\beta_\perp) \Delta_\perp \right] q_\| {b_x} ~, 
\end{align}
where $W_S$ contains the contribution associated with
the perturbed heat fluxes, while $W_{VB}$ comprises the remaining
megneto-kinematic part. This separation provides a convenient framework
for identifying the respective thermal and thermo-mechanical
contributions to the conserved quantity.
Equation (40) shows that an increase in the heat flux parameter $q_\|$ enhances the magnetic field contribution to the invariant, which would otherwise be primarily dominated by the thermal ($W_S$) and parallel velocity ($v_x$) components. At the same time, the form of the invariant reveals its inherently spatially compact character. Uniform or nearly uniform perturbations, for which
\begin{equation}
|\partial_{xx} \, \Phi_i | \, \sim \, |\Delta_\perp \, \Phi_i|  \ll 1 ~,
\end{equation}
do not significantly contribute to the thermo-mechanic structure.
As a result, the perturbation field associated with the invariant is naturally characterized by localized filamentary structures,
corresponding to an aperiodic thermo-mechanic quantity that remains conserved throughout the time 
evolution of the anisotropic flow.

Interestingly, the invariant enforces fixed algebraic relations (``phase'') between perturbation components. In particular, Eq.~(36) shows that a perturbation in the field-aligned velocity $v_x$ must be accompanied by correlated perturbations in the parallel heat flux $s_{\parallel}$ and magnetic field $b_x$, with coefficients determined by the operator $f$, the anisotropy parameter $\beta_{\Delta}$, and the heat-flux amplitude $q_{\parallel}$.
In this sense, the thermo-mechanic invariant defines a \emph{polarization constraint} rather than a dynamical mode: it prescribes how thermal, kinetic, and magnetic perturbations must combine in order to form a time-independent structure.

\section{Discussion and Summary}

In this work we have examined linear perturbations of an incompressible, anisotropic, 
magnetized plasma using an extended 16-moment fluid description with heat fluxes. 
Within this framework, We identify a conserved thermo-mechanic invariant: a time-independent, 
aperiodic perturbation involving heat flux, velocity, and magnetic field fluctuations, 
which does not exist in fluid closures that neglect dynamical heat fluxes. 
Obtained results demonstrate that in addition to the familiar wave-like sector, the 
extended fluid system possesses a distinct stationary (aperiodic) sector governed by 
algebraic constraints.

In the standard CGL framework, where the heat flux degrees of freedom are neglected, no direct analogue of the present invariant is recovered.
Retaining heat flux dynamics qualitatively alters the structure of the linear system: 
the resulting conserved invariant couples the heat-flux, velocity, and magnetic-field degrees of freedom, giving rise to the thermo-mechanic mode.
This behavior highlights that heat fluxes, corresponding to third-order velocity moments of the distribution function, play a dynamically essential role even at the linear level and 
may not be regarded merely as dissipative or transport corrections.

From a physical standpoint, existence of the thermo-mechanic invariant demonstrates that 
third order kinetic moments, represented by parallel and perpendicular heat fluxes, 
can actively participate in the linear dynamics of weakly collisional plasmas. 
In the extended 16-moment formulations, heat fluxes are promoted to autonomous 
dynamical variables whose evolution is coupled to pressure anisotropy, flow, and 
magnetic perturbations.  As a result, heat flux perturbations can enter time invariant 
constructs of linear perturbations rather than merely modifying growth or damping rates 
of individual time dependent modes.

For comparison, Landau-fluid formulations prescribe heat fluxes through closures designed 
to reproduce kinetic damping \cite{Snyder97}. In such models, 
heat flux perturbations are driven by temperature gradients, removing the dynamical freedom 
required for persistent aperiodic solutions. In contrast, the thermo-mechanic invariant 
relies on the autonomous evolution of third-order moments and should not generally 
survive in Landau-fluid closures at larger scales, 
even though kinetic microinstabilities are still expected to regulate pressure anisotropy 
and associated heat flux dynamics at sufficiently small scales \cite{Kunz14,Kunz15,Kunz16}.

Importantly, the conservative invariant found here does not rely on
kinetic resonances or damping mechanisms, but follows from the structural properties 
of the extended 16-moment hierarchy itself, highlighting an importance of qualitative 
extension of anisotropic fluid models beyond the CGL framework.

An important physical implication of this invariant is the emergence of linear memory and 
frozen coordinate in anisotropic plasma dynamics. While other perturbation components may 
propagate or redistribute dynamically, the invariant selectively retains a specific combination 
of thermal, kinetic, and magnetic fluctuations, so that the long time linear response depends 
on specific features of the initial state. Even when oscillatory components decorrelate, the invariant ensures that specific combinations of these fields persist. This provides a mechanism for linear memory in collisionless plasma dynamics. 
This construct may be described as an {\it ``algebraic polarization''}: time-independent algebraic form that fixes relative amplitudes without reference to any wave dynamics. 
Moreover, the invariant is preferentially supported by localized gradients, naturally favoring compact, filamentary aperiodic perturbations.

Although the thermo-mechanic invariant does not correspond to a propagating mode,
this linear memory may nevertheless leave observable imprints in weakly collisional space plasmas.
In particular, the persistence of algebraically constrained combinations of heat flux, 
field aligned velocity, and magnetic field fluctuations implies that certain correlations 
can survive over long times, even as oscillating components strongly decorrelate. 
Such long-lived, low frequency correlations, especially in field aligned fluctuations, 
could provide an observational signature of the aperiodic stationary component identified 
here and offer a complementary perspective on the interpretation of observed variability of 
plasma flows. 

The thermo-mechanical invariant derived here is exact within the incompressible extended fluid description. Restoring compressibility introduces additional degrees of freedom that couple the solenoidal thermo-mechanical dynamics to compressive perturbations and generally destroy the exact conservation law. This behavior is analogous to the Els\"asser invariants (or cross helicity) of incompressible ideal MHD, whose exact conservation relies on the incompressibility and is lost once compressibility is introduced.Still, the importance of such invariants lies in revealing the intrinsic structure of the underlying dynamics.Likewise, the present invariant characterizes the intrinsic structure of the incompressible thermo-mechanical sector.

The loss of the invariant due to compressibility can itself be physically meaningful, although its analysis lies beyond the scope of the present work. Moreover, the invariant may also serve as a useful diagnostic in numerical simulations. Any temporal variation of the invariant provides a direct measure of the coupling between the solenoidal thermo-mechanical and compressive sectors, thereby quantifying compressibility induced departures from the solenoidal dynamics.

A central implication of these results is that the admissible linear dynamics in the 
low frequency regime depend sensitively on the chosen moment closure, a point emphasized in recent discussions of anisotropic plasma turbulence and reduced modeling frameworks \cite{Schekochihin22}. 
By raising the closure by one level in the hierarchy, the extended fluid description introduces 
an additional conservative channel that qualitatively modify the stationary response of the system. 
In contrast, Markovian fluid closures, including Landau-fluid models, suppress autonomous heat flux dynamics and thereby eliminate this stationary sector.

Findings presented in the present paper show that extended 16-moment fluid descriptions 
of collisionless plasmas give rise to conserved structures that significantly alter the 
interpretation of linear dynamics. The thermo-mechanic invariant provides a clear diagnostic 
for identifying persistent aperiodic perturbations and offers a benchmark for assessing the 
fidelity of numerical models at lower frequencies. 
Determining whether this invariant survives nonlinear evolution, and how it may influence 
macroscopic plasma behavior, remains an open question that motivates future investigations 
within extended fluid and kinetic frameworks.

\appendix
\section{Derivation of the invariant generator}
In this Appendix we derive the generator ${\bf g}_i$ of the invariant quantity $W = {\bf g}_i \Phi_i$, starting from the stationarity condition (see Eq. 33):
\begin{equation}
{\bf g}_i {\bf N}_{ik} \Phi_k  = 0 ~.
\end{equation}
which must hold without imposing any particular restriction on the perturbation field. 
For arbitrary nonzero perturbations $\Phi_k \not=0$, a necessary condition for a nontrivial solution is that each column of the operator matrix vanish independently:
\begin{equation}
{\bf g}_i {\bf N}_{ik} = 0 ~,~~ {\rm ~for~each~~} k ~.
\end{equation}
The structure of the matrix operator ${\bf N}_{ik}$ (see Eq. 29) shows that columns $k=2,5$ and $7$ each contain a single nonzero entry. Hence, Eq.~(A1) yields:
\begin{align}
{\bf g}_i {\bf N}_{i2} = {\bf g}_7 {\bf N}_{72} = 0 ~,  \\ 
{\bf g}_i {\bf N}_{i5}  = {\bf g}_3 {\bf N}_{35} = 0 ~, \\
{\bf g}_i {\bf N}_{i7}  = {\bf g}_2 {\bf N}_{27} = 0 ~.  
\end{align}
Since the corresponding elements of the matrix operator ${\bf N}_{ik}$ are non-vanishing, the only solution of Eqs. (A3-5) consistent with arbitrary perturbations is:
\begin{equation}
{\bf g}_2 = {\bf g}_3 = {\bf g}_7 = 0 ~.
\end{equation}
The remaining set of equations from Eq.~(A1) is the following:
\begin{align}
k=1:~~ & -3 q_\parallel {\bf g}_4 \partial_{xx} + {\bf g}_6 \partial_{xx} = 0 ~, \\[2pt]
k=3:~~ & {\bf g}_1 \beta_\perp \Delta_\perp
+ 3 j \beta_\perp {\bf g}_4 \Delta_\perp
- j \beta_\| {\bf g}_5 \partial_{xx} = 0 ~, \\[2pt]
k=4:~~ & 0 = 0 ~, \\[2pt]
k=6:~~ & {\bf g}_1 (\Delta_\perp + \beta_F \partial_{xx})
+ 3 j {\bf g}_4 (\Delta + 2 \beta_\Delta \partial_{xx}) + j \beta_\Delta {\bf g}_5 \partial_{xx} = 0 ~. 
\end{align}
Now Eq. (A6) yields 
\begin{equation}
{\bf g}_6 = 3 q_{\|}{\bf g}_4 ~.
\end{equation}
Hence, by rearranging Eqs.~(A8) and (A10) in terms of spatial derivatives, we obtain:
\begin{align}
{\bf A}_1 \partial_{xx} + {\bf B}_1 \Delta_\perp = 0 ~, \\
{\bf A}_2 \partial_{xx} + {\bf B}_2 \Delta_\perp = 0 ~,
\end{align}
with
\begin{align}
{\bf A}_1 &= -j\beta_\| ~{\bf g}_5 ~, \\
{\bf B}_1 &= \beta_\perp ~({\bf g}_1 + 3j {\bf g}_4) ~, \\
{\bf A}_2 &= (1 + 2 \beta_\Delta) ({\bf g}_1 + 3j {\bf g}_4) + \beta_F {\bf g}_1 - 
j \beta_\Delta ~{\bf g}_5 ~. \\
{\bf B}_2 &= {\bf g}_1 + 3j {\bf g}_4 ~, 
\end{align}
Since $\partial_{xx}$ and $\Delta_\perp$ are independent differential
operators Eq. (A14) requires the minimal operator structure required
to make the two terms compatible: the lowest order choice consistent with
the longitudinal--transverse operator symmetry is to take $A_1 \propto \Delta_\perp$ and $B_1 \propto \partial_{xx}$. Motivated by this fact we assume that
\begin{equation}
{\bf g}_5 \equiv g_0 \, \Delta_\perp ~,
\end{equation}
where $g_0$ is a constant normalization factor. Hence, Eqs. (A12)-(A13) yield the solution as follows:
\begin{equation}
{\bf g}_1 + 3 j {\bf g}_4 = {j g_0 \over \alpha} \partial_{xx} ~,
\end{equation}
\begin{equation}
{\bf g}_1 = {j g_0 \over \alpha \beta_\Delta} \left((1+2 \beta_\Delta) \partial_{xx} + \left( 1+ \alpha \beta_\Delta \right) \Delta_\perp \right) ~.
\end{equation}
Here $g_0$ is an arbitrary normalization factor. Since the invariant
$W$ is determined only up to an overall multiplicative constant, we fix
$g_0$ by choosing:
\begin{equation}
g_0 = -3 \alpha \beta_\Delta ~.
\end{equation}
This particular value is chosen to set ${\bf g}_4 \to \Delta$ in the $\beta_\Delta \to 0$ limit.
Combining the above results, the nonzero components of the generator are
\begin{align}
{\bf g}_1 &= -3j\left[(1+2 \beta_\Delta) \partial_{xx} + (1+\alpha \beta_\Delta)\Delta_\perp \right], \\
{\bf g}_4 &= \beta_F \partial_{xx} + (1+\alpha \beta_\Delta)\Delta_\perp, \\
{\bf g}_5 &= -3\alpha \beta_\Delta \Delta_\perp .
\end{align}
Thus, the solutions obtained from Eqs. (A6), (A11) and (A22)-(A24) provide the
explicit form of the generator components, thereby fully specifying the
structure of the invariant quantity.

\section*{Acknowledgements}

We acknowledge support from the Georgian National Astrophysical Observatory.

\end{document}